\global\def\draftcontrol{0}

   \def\versionno{ gauge theories on ads4 -- draft   }

\catcode`\@=11

\expandafter\ifx\csname draftcontrol\endcsname\relax\global\def\draftcontrol{0}
\fi

{\count255=\time\divide\count255 by 60
\xdef\hourmin{\number\count255}
\multiply\count255 by-60\advance\count255 by\time
\xdef\hourmin{\hourmin:\ifnum\count255<10 0\fi\the\count255}}
\def\draftdate{\number\month/\number\day/\number\year\ \ \ \hourmin }

\newcommand\makepapertitle{\par
  \begingroup
    \renewcommand\thefootnote{\@fnsymbol\c@footnote}%
    \def\@makefnmark{\rlap{\@textsuperscript{\normalfont\@thefnmark}}}%
    \long\def\@makefntext##1{\parindent 1em\noindent
            \hb@xt@1.8em{%
                \hss\@textsuperscript{\normalfont\@thefnmark}}##1}%
     \newpage
     \global\@topnum\z@   
     \@makepapertitle
     \thispagestyle{empty}\@thanks
  \endgroup
  \setcounter{footnote}{0}%
  \global\let\thanks\relax
  \global\let\makepapertitle\relax
  \global\let\@makepapertitle\relax
  \global\let\@thanks\@empty
  \global\let\@author\@empty
  \global\let\@date\@empty
  \global\let\@title\@empty
  \global\let\title\relax
  \global\let\author\relax
  \global\let\date\relax
  \global\let\and\relax
  \def\version{\let\version\@version\@gobble}
}
\def\@makepapertitle{%
  \newpage
   \ifnum\draftcontrol=1 {}
   \version\versionno
   \vskip 3em%
   \else
   \hfill\hbox to 3cm {\parbox{4cm}{\@pubnum}\hss}%
   \vskip 3em%
   \fi
   \begin{center}%
   \let \footnote \thanks
     {\LARGE {\@title}}%
     \vskip 1.5em%
     {\normalsize
       \lineskip .5em%
       \begin{tabular}[t]{c}%
         \@author
       \end{tabular}\par}%
     \vskip 1.5em%
     {\@bstract}%
     \end{center}%
     \vskip 1.5em
     \@date%
   \par
}

\gdef\@pubnum{}
\def\pubnum#1{%
  \gdef\@pubnum{#1}}

\gdef\@bstract{}
\def\Abstract#1{%
  \gdef\@bstract{%
   \parbox{\textwidth-0pc}{%
   \centerline{\bf Abstract}\penalty1000%
\kern.2cm%
\noindent
\renewcommand\baselinestretch{1.0}%
{#1}}}
}

\def\ps@paper{\let\@mkboth\@gobbletwo%
     \ifnum\draftcontrol=1
	\def\@oddfoot{\hbox to \textwidth{\tiny \versionno \hfil\tiny\draftdate}%
	\hskip -\textwidth \hbox to \textwidth{\hfil\rm\thepage\hfil}}%
     \else\def\@oddfoot{\hbox to \textwidth{\hfil\rm\thepage\hfil}}
     \fi
     \let\@evenfoot\@oddfoot
}

\def\body{\clearpage
          \pagestyle{paper}
	}

\def\@version#1{\ifnum\draftcontrol=1
\typeout{}\typeout{#1}\typeout{}
\vskip3mm\centerline{\hbox{\fbox{\normalsize{\tt DRAFT -- #1 -- }
                   {\draftdate}}}}\vskip3mm
\fi}
\let\version\@version
\long\def\eqlabel#1{\ifnum\draftcontrol=1
                    \tag@false  
                    \tag*{(\theequation) \hbox to -0.2cm{\hspace{0cm}\small{#1}\hss}}
                    \refstepcounter{equation}
                    \edef\@currentlabel{\theequation}
                    \ltx@label{#1}          
                    \else
                    \label{#1}
                    \fi
                    }
\let\st@bibitem\@bibitem
\let\st@lbibitem\@lbibitem
\ifnum\draftcontrol=1
  \def\@bibitem#1{%
    \st@bibitem{#1}\a@@label{#1}\ignorespaces}
  \def\@lbibitem[#1]#2{%
    \st@lbibitem[#1]{#2}\a@@label{#2}\ignorespaces}
  \def\a@@label#1{%
    \gdef\a@lab{\smash{\normalfont\small#1}}
    \ifvmode
      \if@inlabel
        \global\setbox\@labels\hbox{%
          \llap{\a@lab\let\a@lab\relax
                \kern\@totalleftmargin\kern\marginparsep}%
          \box\@labels}%
      \fi
    \fi}
\fi

\documentclass[12pt,letterpaper]{article}

\usepackage{amsmath,amssymb,array,calc,rotating,epsfig,psfrag}
\usepackage[nosort]{cite}

\ifnum\draftcontrol=1
\fi

\renewcommand\baselinestretch{1.25}
\renewcommand\section{\@startsection {section}{1}{\z@}%
                                   {-3.5ex \@plus -1ex \@minus -.2ex}%
                                   {2.3ex \@plus.2ex}%
                                   {\normalfont\large\bfseries}}
\renewcommand\subsection{\@startsection{subsection}{2}{\z@}%
                                   {-3.25ex\@plus -1ex \@minus -.2ex}%
                                   {1.5ex \@plus .2ex}%
                                   {\normalfont\normalsize\bfseries}}
\renewcommand\subsubsection{\@startsection{subsubsection}{3}{\z@}%
                                   {-3.25ex\@plus -1ex \@minus -.2ex}%
                                   {1.5ex \@plus .2ex}%
                                   {\normalfont\normalsize\it}}
\renewcommand\paragraph{\@startsection{paragraph}{4}{\z@}%
                                   {-3.25ex\@plus -1ex \@minus -.2ex}%
                                   {1.5ex \@plus .2ex}%
                                   {\normalfont\normalsize\bf}}

\numberwithin{equation}{section}

\def\ie{{\it i.e.}}

\def\revise#1       {\raisebox{-0em}{\rule{3pt}{1em}}%
                     \marginpar{\raisebox{.5em}{\vrule width3pt\
                     \vrule width0pt height 0pt depth0.5em
                     \hbox to 0cm{\hspace{0cm}{%
                     \parbox[t]{4em}{\raggedright\footnotesize{#1}}}\hss}}}}

\newcommand\nxt[1]  {\\\fnxt#1}

\def\calb         {{\cal B}}

\def\calf         {{\cal F}}
\def\calg         {{\cal G}}

\def\call         {{\cal L}}
\def\calm         {{\cal M}}
\def\caln         {{\cal N}}
\def\calo         {{\cal O}}
\def\calp         {{\cal P}}

\def\calv         {{\cal V}}

\def\del          {\partial}

\def\Im           {{\rm Im\hskip0.1em}}

\def\sqr#1#2{{\vcenter{\vbox{\hrule height.#2pt
 \hbox{\vrule width.#2pt height#1pt \kern#1pt
 \vrule width.#2pt}\hrule height.#2pt}}}}

\newcommand{\ft}[2]{{\textstyle{\frac{#1}{#2}}}}

\def\om{\Omega}

\def\a{\alpha}

\def\r{\rho}

\def\dd{\delta}
\def\tg{\tilde{g}}
\def\LL{\Lambda}
\def\ww{\omega}
\def\n{\nabla}
\def\tn{\tilde{\nabla}}
\def\bG{{\bar{G}}}

\def\tcalm{\tilde{\calm}}
\def\bD3{\overline{D3}}

\def\ss{\Sigma}

\catcode`\@=12

\begin{document}


\title{Gauge theories on hyperbolic spaces\\
and 
dual wormhole instabilities 
}

\pubnum{%
hep-th/0402174}
\date{February 2004}

\author{
Alex Buchel\\[0.4cm]
\it Department of Applied Mathematics\\
\it University of Western Ontario\\
\it London, Ontario N6A 5B7, Canada\\
\it Perimeter Institute for Theoretical Physics\\
\it Waterloo, Ontario N2J 2W9, Canada\\[0.2cm]
}

\Abstract{
We study supergravity duals of strongly coupled four dimensional gauge
theories formulated on compact quotients of hyperbolic spaces. The
resulting background geometries are represented by Euclidean
wormholes, which complicates establishing the precise gauge
theory/string theory correspondence dictionary.  These backgrounds
suffer from the non-perturbative instabilities arising from the
$D3\bD3$ pair production in the background four-form potential. We
discuss conditions for suppressing this Schwinger-like instability.
We find that Euclidean wormholes arising in this construction develop
a naked singularity, before they can be stabilized.
}


\makepapertitle

\body

\version\versionno

\section{Introduction}
Diffeomorphism invariance of a gravitational theory implies that classical 
backgrounds related by  coordinate transformations are physically equivalent. 
This is no longer the case once quantum effects are taken into account. The reason is 
simply because different space-like foliations of the background geometry lead to  different 
definitions of a time ( and thus a Hamiltonian) of a quantum system.   
This has profound implications for the gauge theory/string theory 
correspondence\footnote{For a review see \cite{adscft}.} 
\cite{m97}. In the simplest case, the holographic correspondence of Maldacena 
relates $\caln=4$ $SU(N)$ supersymmetric Yang-Mills (SYM) theory in $R^{3,1}$
and type IIB supergravity in $AdS_5\times S^5$, where $AdS_5$ is written in 
Poincare' patch coordinates. As emphasized in \cite{bt}, even though classical (Euclidean) 
$AdS_5$ foliations\footnote{Corresponding Lorentzian foliations have 
$R^{3,1},\ dS_4,\ AdS_4$ slices.} by $R^4,\ S^4,\ H_4$ are related by 
coordinate transformations, the corresponding gauge theories are physically inequivalent. 
This is so because a {\it classical} supergravity background (in the large $N$ limit, and for the 
large 't Hooft gauge theory coupling) is equivalent to the full {\it quantum} gauge theory 
on the corresponding slices. The correspondence between gauge theories on curved space-times
and gravitational duals becomes more involved for nonconformal gauge theories \cite{bt,bo1,bo2,bo3,b2,bgh,ml}.    

Quite intriguing, certain supergravity backgrounds holographic to gauge theories on 
negatively curved space-times are represented by wormhole solutions \cite{bgh,ml}.
As stressed in \cite{ml}, existence of multiple boundaries in these Euclidean 
supergravity solutions makes it difficult to establish a detailed dictionary for the 
gauge/string theory correspondence. Moreover, the negative curvature of the supergravity 
boundary leads to a nonperturbative instability\footnote{Strictly speaking, 
the instability exists only for a compact negatively curved boundary.} 
due to the $D3\bD3$ pair production \cite{sw}. 
The latter suggests that wormhole solutions arising in this construction are somewhat 
unphysical, and should disappear once nonperturbative instabilities 
are removed. In this paper we study nonperturbative instabilities of strongly coupled four 
dimensional gauge theories
on smooth compact quotients of hyperbolic spaces, and existence of non-perturbatively stable Euclidean 
supergravity wormholes  representing their holographic dual.

Since instabilities on the supergravity side are associated with
the tachyonic modes of the dual gauge theory, the natural way to
eliminate them is to remove tachyons from the gauge theory spectrum. 
The gauge theory tachyons come from  conformally  
coupled scalars, which were massless prior to introducing background
space-time curvature. Indeed, an effective potential for such a scalar 
$\phi$ is\footnote{We assume $\phi$ to be canonically normalized, 
\ie, it has a kinetic term $-\ft 12 (\del\phi)^2$.} 
\begin{equation}
V_{\phi}=\frac {1}{12}\ R_4\ \phi^2\,,
\eqlabel{tach1}
\end{equation}
giving rise to a negative mass square $m^2_{\phi}=\ft 16 R_4$ for 
negative 4d Ricci scalar curvature $R_4$. These scalars are 'true'
tachyons only when the spacial directions of the gauge theory 
background are compactified. This is also the case
with the instabilities on the supergravity side: they are present 
only for compact spatial directions. In the noncompact case,
say $H_4$ of radius $L$, the mass of a conformally coupled scalar is above 
the Breitenlohner-Freedman bound
\begin{equation}
m^2_{\phi}=-\frac{2}{L^2}\qquad  >\qquad  m_{BF}^2=-\frac{9}{4L^2}\,,
\eqlabel{mBF}
\end{equation}   
and thus does not lead to any instabilities. Similarly, 
in this case the potential barrier to create a $D3\bD3$ pair 
in the dual supergravity background is infinite, simply because 
$H_4$ volume is infinite. For this reason, 
we consider four-dimensional gauge theories
on  $S^1\times
\ss_3$ and $\ss_4$, where $\ss_n$ is a smooth, compact, finite volume  quotient of a
hyperbolic space $H_n$ by a discrete subgroup $\Gamma$ of its 
$SO(n,1)$ symmetry group, $\ss_n=H_n/\Gamma$.  

In the next section we study $D3$ probe brane dynamics in 
supergravity dual to $\caln=4$ $SU(N)$ SYM theory on $S^1\times
\ss_3$, which  is a Euclidean continuation of this gauge theory on $R\times
\ss_3$ at finite temperature.
The motivation to study this potential mechanism for lifting 
tachyonic modes comes from  finite temperature field
theory intuition: there,  a thermal mass can be induced to lift otherwise 
tachyonic mode. We find that the instability still
persists. In fact, no thermal mass is induced for the conformally
coupled scalar in the regime relevant for the instability. 
We speculate as to why this happens. 
A natural way to lift a tachyon is to give it a bare 
mass.\footnote{This mechanism of stabilization of supergravity 
backgrounds dual to gauge theories on $\ss_n$ was also suggested in
\cite{ml}.} On the dual supergravity side, this corresponds to turning
on 3-form fluxes (for fermionic masses), and/or deforming 
the asymptotic background geometry (for bosonic masses).      
In section 3, we study a  $D3$ probe brane dynamics in a general
warped type IIB background with fluxes. We present a rather simple equation 
for the probe brane effective potential, and obtain some universal results      
concerning non-perturbative $D3\bD3$ pair production instability. 
In section 4, we study in details the supergravity dual to 
$\caln=2^*$ $SU(N)$ SYM theory on $\ss_4$. In Minkowski space, the 
notation '$\caln=2^*$' means that the theory is  obtained 
from the parent $\caln=4$ SYM theory by giving the same  mass to two $\caln=1$
chiral multiplets (a mass to $\caln=2$ hypermultiplet). We will keep the same 
label for the gauge theory, even though our deformation completely breaks supersymmetry. 
In fact, we will discuss massive $\caln=2^*$ supergravity renormalization group (RG) 
flows\footnote{$\caln=2^*$ supergravity RG flows on $R^{3,1}$ were constructed in \cite{pw} (PW). 
Deformations of the PW solution closely related to the topic of this paper were constructed in \cite{b2}.
As in \cite{pw,b2}, $\caln=2^*$ flows discussed here admit an exact, analytical 
lift to a complete ten-dimensional type IIB supergravity background.} on $\ss_4$
induced by (generically) different masses for the bosonic and fermionic 
components of the $\caln=2$ hypermultiplet. Pertaining to this RG flow we obtain the following results.
\nxt  
Despite the fact that we turn on masses for bosonic and fermionic components 
for the $\caln=2$ hypermultiplet only, and thus leaving  the chiral multiplet in the $\caln=2$ 
vector multiplet massless, it is possible to remove all tachyonic instabilities from the 
$D3$ probe brane effective action. This eliminates  catastrophic instability of the 
supergravity background associated with $D3\bD3$  pair production. Interestingly,
to achieve the latter, one necessarily have to turn on different bare masses for the
bosonic and fermionic components of the hypermultiplet. For equal bosonic and fermionic 
masses, the tachyonic instability of the $D3$ probe is  {\it identical} to the 
instability with zero masses, \ie, for the supergravity background dual to $\caln=4$ 
gauge theory on $\ss_4$.
\nxt
For zero masses of the hypermultiplet components, the dual  supergravity background represents the simplest 
Euclidean wormhole solution \cite{ml}:
\begin{equation}
ds_{10}^2=\biggl[L^2\cosh^2\left(\frac rL\right)\ \left(d\ss_4\right)^2+dr^2\bigg]+L^2\ \left(dS^5\right)^2\,,
\eqlabel{wh1}
\end{equation}   
where the metric in $[\cdots]$ is that of the $AdS_5/\Gamma$ of radius $L$ with $\ss_4=H_4/\Gamma$ foliations, 
and $(dS^5)^2$ is the metric of the round $S^5$ of unit radius. We analytically construct  
deformations of this wormhole solution to leading order in bosonic and fermionic masses of the 
hypermultiplet components. The deformed geometry is still a smooth wormhole solution.    
\nxt
We study numerically the mass-deformed wormhole \eqref{wh1} as we increase the mass of the 
fermionic  components of the hypermultiplet, $m_f$. For simplicity, we keep vanishing  the  mass of the bosonic 
components of the hypermultiplet, as well as the 
vacuum expectation values for fermionic and bosonic bilinear condensates. The tachyonic 
instabilities in the $D3$ probe brane effective action are removed provided  
\begin{equation}
m_f^2\ \ge\  m_{critical}^2=\frac{12}{L^2}\,.
\eqlabel{mcrit}
\end{equation}
However, well before we reach the critical mass in \eqref{mcrit}, 
the geometry develops a naked singularity. For ultraviolet initial conditions 
for the RG flow as above, this happens for $m_f\ge m_{singular}$, where 
\begin{equation}
\frac{m_{singular}}{m_{critical}}\approx 0.3719\cdots\,.
\eqlabel{msing}
\end{equation}

Finally, we would like to point out that though we apply the effective potential for a
$D3$ probe brane of section 3 to study instabilities of the gauge theories on negatively curved space-times, 
the equations for the effective potential \eqref{dpotg}, \eqref{dpotg0} are valid for any 
sign of the  gauge theory background  cosmological constant. As we briefly mention in section 
3, this observation provides a simple explanation for the large $\eta$-parameter 
for the $D3$-brane inflation in the Klebanov-Strassler \cite{ks} throat geometries, 
presented in \cite{br}. We expect that \eqref{dpotg}, \eqref{dpotg0} will be useful 
in search of single-field slow-roll brane inflationary models in type IIB supergravity, and propose a brane 
inflationary model with small $\eta$.

\section{$\caln=4$ SYM on $R\times \ss_3$ at finite temperature}

Consider the nonextremal deformation of the $AdS_5/\Gamma\times S^5$ 
solution, where $AdS_5/\Gamma$ is foliated with 
$R\times \ss_3$. Here,  $\ss_3=H_3/\Gamma$ is a smooth, compact, finite volume   
quotient of the three dimensional hyperbolic space $H_3$ by a discrete 
subgroup of its $SO(3,1)$ symmetry group\footnote{
The relevant background was constructed in\cite{em}.}. 
Following \cite{m97}, we want to interpret this as a supergravity dual to strongly coupled 
$\caln=4$ $SU(N)$ 
gauge theory on $R\times \ss_3$ at finite temperature. 
As usual, after the analytical continuation, the gauge theory background geometry becomes $S^1\times \ss_3$,
where the euclidean time periodicity coincides with inverse temperature, $t_E\sim t_E+1/T$. 
After reviewing the properties of the dual supergravity 
background, we study the dynamics of $D3, \bD3$ probes. We find that a $\bD3$ brane is stabilized 
at the origin of the Euclidean supergravity background\footnote{The point 
where $S^1$ shrinks to zero size.}, with vanishing action. On the other hand, 
the effective action of a $D3$ brane is unbounded from below. This instability comes from the
tachyonic mode of a $D3$ probe brane effective action, originating from the 
conformally coupled scalar corresponding to moving a brane in a radial direction. 
In what follows we  refer to this (canonically normalized) 
scalar as a {\it radion}, $\phi$. 
We find that while near the origin the effective radion mass squared $m_{rad}^2$ is positive,
\begin{equation}
\begin{split}
m_{rad}^2=&2\pi^2 T^2\,,\qquad \phi^2\ll \frac{g_{YM}^2
N}{\r^2}\,,\qquad \r T \gg 1\,,\\
m_{rad}^2=&\frac{1}{2\r^2}\,,\qquad \phi^2\ll \frac{g_{YM}^2
N}{\r^2}\,,\qquad 0\le\r (T-T_0)\ll 1\,,\qquad \r T_0=\frac {1}{2\pi}\,,
\end{split}
\eqlabel{mr0}
\end{equation}  
( $g_{YM}^2 N$ is the gauge theory 't Hooft coupling) it becomes tachyonic close to the boundary 
\begin{equation}
m_{rad}^2=-\frac{1}{\r^2}\,,\qquad \phi^2\gg \frac{g_{YM}^2 N}{\r^2}\,,
\eqlabel{mri}
\end{equation}
as appropriate for the conformally coupled scalar on $S^1\times \ss_3$, \eqref{tach1}, with $\ss_3$ 
radius of curvature $\r$, 
\begin{equation}
R_{S^1\times \ss_3}=3\cdot \left(-\frac {2}{\r^2}\right)\,.
\eqlabel{rcurv}
\end{equation} 
Notice that \eqref{mri} is independent of the temperature, for which we provide a heuristic physical 
explanation later in the section. 
Because of the unbounded character of a $D3$ brane action close to the 
boundary (in the regime \eqref{mri}), and the fact that 
a barrier to create a $D3\bD3$ pair is finite (it is of order $\ft 1N\sim \ft {1}{g_s}$),
it is always energetically favorable to create $D3\bD3$ pairs near the boundary. Once created, 
a $D3$ brane will move to the boundary, while $\bD3$ will move into the bulk. Such a  process 
reduces the free energy of the gravitational background, and it's four-form 'charge'.
In  this sense it is very similar to the Schwinger mechanics for the electron-positron 
pair production in strong electric field. Eq.~\ref{mri} implies that finite temperature 
can not eliminate this non-perturbative instability.

The gravitational background considered in this section 
is not a wormhole. Explicit wormhole example based on a gravitational dual to 
Euclidean gauge theory on $\ss_4$ is discussed in section 4. Nonetheless, 
the physics of that wormhole  instability is the same as discussed above. This is so, because the 
$D3\bD3$ pair-production instability near a negatively curved boundary is a local phenomenon,
and thus is insensitive to the presence of multiple boundaries.

\subsection{The dual supergravity background}

For the dual supergravity background we take the following metric ansatz 
\begin{equation}
ds_{10}^2=-c_1^2 \left(dt\right)^2+c_2^2 \left(d\ss_3\right)^2
+c_3^2 \left(dr\right)^2 +c_4^2\left(dS^5\right)^2\,,
\eqlabel{m10}
\end{equation}
where $c_i=c_i(r)$ and $\left(d\ss_3\right)^2$ and
 $\left(dS^5\right)^2$ are the metrics on the 'unit radius of curvature' $\ss_3$ and $S^5$ 
correspondingly. Additionally, there is a five form flux, that we take
 to be of the form 
\begin{equation}
F_5=\calf_5+\star\calf_5\,,\qquad \calf_5=-4 L^4 d{\rm vol}_{S^5}\,.
\eqlabel{5form}
\end{equation}
Solving type IIB supergravity equations of motion we find the following solution \cite{em}
\begin{equation}
\begin{split}
&c_1=f^{1/2}\,,\qquad c_2=\frac{\rho r}{L}\,,\qquad c_3=\frac {1}{f^{1/2}}\,,\qquad c_4=L\,,\\
&f=\frac {r^2}{L^2}-\frac{L^2}{\rho^2}-\frac{\mu}{r^2}\,,
\end{split}
\eqlabel{sol}
\end{equation}
where $\rho$ is the 'radius of curvature of the gauge theory'
hyperbolic three-space,
$\mu$ is the nonextremality parameter.  
The thermodynamics of this  black hole  was studied in details by Emparan \cite{em}, 
where it was found that the specific heat is always positive. 
This result is somewhat surprising, as we would 
expect that the gauge theory instabilities would show up as 
thermodynamic instabilities, \cite{gm}. 

For later references we present the expression for the black hole \eqref{sol} temperature
\begin{equation}
T=\frac{1}{2\pi r_0}\ \left(\frac{r_0^2}{L^2}+\frac{\mu}{r_0^2}\right)\equiv \frac{1}{2\pi \r b_0^3 L^3}
\left(b_0^4L^2+\mu\r^4\right)\,, 
\eqlabel{htem}
\end{equation}
where  $r_0\equiv b_0 L/\r$ is the position of the horizon (the largest root of)
\begin{equation}
f(r_0)=0\qquad \Longleftrightarrow \qquad b_0^4 L^2-b_0^2 L^4-\mu\r^4=0\,.
\eqlabel{r0}
\end{equation}

\subsection{Probe dynamics}
\subsubsection{$D3$ brane}

Let's consider a  D3 probe dynamics in above geometry.
We consider the case when the probe moves in a radial ($r$) direction only, $r_1=r_1(t)$. 
Dependence of $r_1$ on the coordinates of $\ss_3$ does not 
modify the story in any substantial way (there is a slight
modification though because $c_2\ne c_1$)
The probe action reads \cite{pol2}
\begin{equation}
S_{D3}=-T_3\int_{R\times \ss_3}d^4\xi\sqrt{-g(r_1)}+T_3\int_{R\times \ss_3}C^{(4)}(r_1)\,,
\end{equation}
where $T_3$ is a three-brane tension, and $C^{(4)}$ is a four-form potential 
giving rise to the five-form flux \eqref{5form}.
As the radion $r_1$ changes with time slowly, we find 
\begin{equation}
S_{D3}=\int_{R\times \ss_3}dt\ \r^3dvol_{\ss_3}\ \left(\ft 12 T_3 c_1^{-1}\left(\frac{c_2}{\r}\right)^3c_3^2
(\del_t r_1)^2-\calv(r_1)
\right)\,,
\eqlabel{effac}
\end{equation}
where $\calv(r_1)$ is the radion potential energy
\begin{equation}
\calv(r_1)=\frac{T_3}{\r^3}\ \biggl(c_1 c_2^3-C^{(4)}\biggr)\,.
\eqlabel{pot}
\end{equation} 
Canonical normalization of the scalar $r_1
\to \phi$, is achieved with  
\begin{equation}
T_3 c_1^{-1}\left(\frac{c_2}{\r}\right)^3c_3^2
(\del_t r_1)^2\equiv (\del_t \phi)^2\,.
\eqlabel{cannom}
\end{equation} 
We then get the 'physical' potential energy $\calv_{rad}$ for 
large $\phi$
\begin{equation}
\calv(r_1)=\calv_{rad}(\phi)=-\frac{\phi^2}{2\rho^2}-\frac{T_3}{2L^2}\biggl(\mu+\frac
{7L^6}{4\rho^4}\biggr)+\calo(\phi^{-2})\,.
\eqlabel{potph}
\end{equation}
resulting in the radion mass \eqref{mri}.

Eq.\eqref{potph} gives the radion potential $\calv_{rad}(\phi)$ for  large $\phi$.
For completeness, we also present   expressions for  $\calv_{rad}$ near the black hole horizon
(or the origin of the corresponding Euclidean geometry). This can be best done by using the 
canonically normalized radion, defined by \eqref{cannom}, as a radial coordinate for the 
background \eqref{m10}. One can then solve the equations of motion for $c_i$ in this radial gauge 
as power series. Using the  following boundary condition 
(this can always be done)
\begin{equation}
\phi\bigg|_{horizon}=0
\eqlabel{bcond}
\end{equation}
near the horizon (small $\phi$), we find the following expansion
\begin{equation}
\begin{split}
 \calv_{rad}(\phi)=&\frac{T_3}{\r^3}\ b_0^3 a_0\  \phi^2-
\frac {T_3^{1/2}}{\r^{3/2} L} b_0^{3/2} a_0^{3/2}\ \phi^4+ 
a_0^2 \frac{10 b_0^2-3 L^2}{20 b_0^2  L^2}\ \phi^6-\frac{3
a_0^{5/2} \r^{3/2}}{40 T_3^{1/2} L b_0^{7/2}}\ \phi^8+\calo(\phi^{10})\,,
\end{split}
\eqlabel{pothor}
\end{equation}
 where 
\begin{equation}
\begin{split}
a_0=&\frac{\r(2b_0^2-L^2)(b_0^4 L^2+\mu \rho^4)}{4 T_3 b_0^7 L^4}\,,\\
0=&b_0^4 L^2-b_0^2 L^4-\mu\rho^4\,,\qquad b_0^2 \ge  L^2\,. 
\end{split}
\eqlabel{a0b0}
\end{equation}
Consider first the high temperature limit, so that
\begin{equation}
\r T\gg 1\,.
\eqlabel{hight}
\end{equation}
Then, 
\begin{equation}
\begin{split}
T=&\frac{b_0}{\pi\r L}\left(1+\calo\left(\frac{L}{b_0}\right)\right)\,,\qquad \mu=\frac{b_0^4 L^2}{\r^4}
\left(1+\calo\left(\frac{L}{b_0}\right)\right)\,,\\
a_0=&\frac{\r}{T_3 b_0 L^2}\left(1+\calo\left(\frac{L}{b_0}\right)\right)\,,\qquad b_0\gg L\,,
\end{split}
\eqlabel{highTex}
\end{equation}
and 
\begin{equation}
\begin{split}
\calv_{rad}(\phi)=&\frac{b_0^2}{L^2\r^2}\ \phi^2-\frac{1}{T_3 L^4}\ \phi^4+\calo\left(\frac{\phi^6\r^2}
{b_0^2L^6 T_3^2}\right)\\
=&\pi^2 T^2\ \phi^2-\frac{1}{T_3 L^4}\ \phi^4
+\calo\left(\frac{\phi^6}{T^2 L^8 T_3^2}\right)\,,\qquad \r T\gg1\,.
\end{split}
\end{equation}
Noting that $T_3 L^4=\ft{g_s N}{2\pi^2}=\ft{g_{YM}^2 N}{2\pi^2}$, we obtain the 
effective radion mass as in \eqref{mr0}.  
In  the low temperature limit\footnote{Here, by low temperature we mean the $\mu\to 0_+$ limit
of the nonextremality parameter. As discussed in details in \cite{em}, the 
black hole \eqref{sol} has a nonzero temperature (and horizon area) 
at $\mu=0$. It exists also for $0>\mu \ge\mu_{extremal}=-\frac{L^6}{4\r^4}$.}, 
\begin{equation}
0\le\r (T-T_0)\ll 1\,,\qquad \r T_0=\frac {1}{2\pi}\,,
\eqlabel{low}
\end{equation}
we find 
\begin{equation}
\begin{split}
T\approx T_0=&\frac{1}{2\pi\r }\,,\qquad \mu\approx 0\,,\\
a_0\approx \frac{\r}{4L^3 T_3}&\,,\qquad \frac{b_0-L}{L}\ll 1\,,
\end{split}
\eqlabel{lowTex}
\end{equation}
and 
\begin{equation}
\begin{split}
\calv_{rad}(\phi)=&\frac{1}{4\r^2}\ \phi^2-\frac{1}{8 T_3 L^4}\ \phi^4+\calo\left(\frac{\phi^6\r^2}
{L^8 T_3^2}\right)\\
=&\pi^2 T_0^2\ \phi^2-\frac{1}{8T_3 L^4}\ \phi^4
+\calo\left(\frac{\phi^6}{T_0^2 L^8 T_3^2}\right)\,,\qquad 0\le \r (T-T_0)\ll 1\,.
\end{split}
\eqlabel{vlowt}
\end{equation}

\subsubsection{$\bD3$ brane}
Similar analysis can be done for a $\bD3$ brane probe. 
Here, for large $\phi^2\r^2\gg T_3 L^4$ its effective potential is
\begin{equation}
\calv_{\bD3}(\phi)=\frac{2}{T_3 L^4}\ \phi^4+\frac{11}{2\rho^2}\ \phi^2+\calo(\phi^0)\,,
\eqlabel{potbD3}
\end{equation}
while for   $\phi^2\r^2\ll T_3 L^4$, we have 
\begin{equation}
\begin{split}
 \calv_{\bD3}(\phi)=&\frac{T_3}{\r^3}\ b_0^3 a_0\  \phi^2+
\frac {T_3^{1/2}}{\r^{3/2} L} b_0^{3/2} a_0^{3/2}\ \phi^4+ 
a_0^2 \frac{10 b_0^2-3 L^2}{20 b_0^2  L^2}\ \phi^6+\frac{3
a_0^{5/2} \r^{3/2}}{40 T_3^{1/2} L b_0^{7/2}}\ \phi^8+\calo(\phi^{10})\,.
\end{split}
\eqlabel{bD3pothor}
\end{equation}
Thus, a $\bD3$ experiences an attractive potential and is pulled away from the boundary. 
Upon analytical continuation, the Euclidean time direction is compactified 
with periodicity $1/T$. This Euclidean-time circle shrinks to zero size at $\phi=0$,
precisely where $\bD3$  is stabilized. $\bD3$ Euclidean action will thus vanish at
$\phi=0$.

\subsection{Thermal mass for the $D3$ brane radion?}
In previous section we found that no thermal mass is generated for a $D3$ probe brane radion close to the 
boundary. On the  other hand,  the effective mass of a    $\bD3$  brane radion near the 
boundary differs from that of the boundary conformally coupled scalar (it has even a wrong sign), 
though it is still temperature independent. We do not have a field-theoretical explanation 
for this. It could very well be a strong coupling effect, and thus inaccessible to the perturbative reasoning.
Nonetheless, it is tempting to draw an analogy to finite temperature four-dimensional 
scalar field theory with quartic self-coupling. There, starting with a zero temperature 
symmetry breaking potential
\begin{equation}
V(\phi,T=0)=-\frac{\mu^2}{2}\ \phi^2+\frac{\lambda}{4}\ \phi^4\,,
\end{equation}  
one finds that interactions with a high temperature thermal background introduce  corrections
\begin{equation}
V(\phi,T)=V(\phi,0)+\frac{T^2}{24}\ V''(\phi,0)+\cdots\,, 
\eqlabel{potfint}
\end{equation} 
where the derivatives are with respect to $\phi$.
As a result, for $\lambda>0$ and sufficient large temperature, the effective 
mass square of the scalar field at the origin ($\phi=0$) can become positive
\begin{equation}
m^2_{\phi}=\frac{\lambda}{4}\ T^2-\mu^2>0\,,\qquad T^2>\frac{4\mu}{\lambda}\,.
\eqlabel{phaset}
\end{equation}
Precisely this mechanism for lifting the non-perturbative instability of the supergravity 
dual to $\caln=4$ gauge theory on $R\times \ss_3$ we had in mind earlier in this section. 
The likely reason why it did not work, is because the $D3$ brane radion near the boundary 
(where it is tachyonic) does not have a quartic self-coupling: the Laurent power series 
expansion of it's effective potential starts with a $\calo(\phi^2)$ term,  \eqref{potph}, as the $\calo(\phi^4)$ 
term vanishes due to the asymptotic supersymmetry. 
Alternatively, while the tachyonic contribution to the $D3$ brane radion mass (due to the curvature coupling) 
is classical, the thermally induced mass correction is radiative. Radiative effective potential 
corrections typically flatten out for large values of the field VEV. Thus, for 
large values of $\phi$ they can not counteract classical tachyonic curvature induced mass\footnote{Strictly 
speaking, effective potential \eqref{potfint}, leading to \eqref{phaset}, is valid only near the origin in the field space.}. 
Perturbative analysis indicating such  saturation of the thermally induced mass in 
finite-temperature $\phi^4$-theory was reported in \cite{hk}.

\section{Probe branes in  generic flux backgrounds}

Having failed to eliminate nonperturbative instability due to $D3\bD3$ 
pair production in supergravity duals to gauge theories on $\ss_3$ with finite temperature, 
we now turn to a more mundane method:  we  give gauge theory would-be  tachyons 
sufficiently large bare mass. On the supergravity side, this is mapped into turning on 
appropriate three-form fluxes. This leads us to study $D3, \bD3$ probe brane dynamics 
in general warped geometries with fluxes. Curiously, one can obtain a rather 
simple equation for the effective probe brane potential. Our discussion 
is rather general, in particular, we do not specify the sign of the curvature of a 
four-dimensional slice wrapped by a $D$-brane. 
We explain under what conditions  fluxes can 'lift' the $D3$ brane radion 
close to the negatively curved boundary. In section 4, this idea will be 
explicitly implemented for $\caln=2^*$ PW flow on $\ss_4$. 
Additionally, we comment on the utility of \eqref{dpotg}, \eqref{dpotg0} 
for the cosmological brane inflationary model building.

\subsection{$D3, \bD3$ probe dynamics in warped geometries with fluxes}

Consider a generic type IIB supergravity flux background on direct warped 
product $\calm_4\times \tcalm_6$. Specifically, we take the metric ansatz (in Einstein 
frame) to be 
\begin{equation}
\begin{split}
ds_{10}^2&=e^{2 A(y)}\  ds^2_{\calm_4}(x)+e^{-2 A(y)}\   
ds^2_{\tilde{\calm}_6}(y)\\
&=e^{2 A(y)}\ g_{\mu\nu}(x) dx^{\mu}dx^{\nu} 
+ e^{-2 A(y)} \tg_{mn}(y) dy^{m}dy^{n}\,,
\end{split}
\eqlabel{10metricgg}
\end{equation}
where $\calm_4$ is taken to be a smooth compact  Einstein manifold, \ie,
\begin{equation}
r_{\mu\nu}^{(4)}(x)=\LL\ g_{\mu\nu}(x)\,,
\eqlabel{4g}
\end{equation}
and $\tilde{\calm}_6$ is a six dimensional non-compact manifold. 
The four-dimensional cosmological constant $\LL$ can be of either sign (or zero).
Additionally we assume that all fluxes, dilaton depend 
on $y$ only.  For the 5-form $\calf_5$ we assume 
\begin{equation}
\calf_5=\left(1+\star\right) \left[d\ww\wedge vol_{\calm_4}\right]\,,  
\eqlabel{5formg}
\end{equation}
where $vol_{\calm_4}$ is the volume form on $\calm_4$.
The complex 3-form flux $G=G(y)$ is transverse to $\calm_4$, also the 
type IIB axiodilaton (in convention of \cite{schwarz83}) $\tau=\tau(y)$ satisfies 
\begin{equation}
\tau\equiv C_{(0)}+i e^{-\Phi}=i\frac{1+\calb}{1-\calb}\,.
\eqlabel{axidil}
\end{equation}
Equations of motion for these warped geometries in the case of $\LL=0$ were 
derived in \cite{gkp}, and for  general Einstein manifolds $\calm_4$ in \cite{b1}:
\begin{equation}
\begin{split}
\tn^2 e^{4A}=& 
\ft {1}{12} e^{8A} G\bG +e^{-4 A} \biggl(
16\ 
\left(\tn\ww\right)^2+\left(\tn e^{4A}\right)^2 
\biggr)+4\LL\,,
\end{split}
\eqlabel{aeq}
\end{equation}
\begin{equation}
\begin{split}
\tn^2\ww=&
2 e^{-4A}\tn\ww \tn e^{4A}+
\ft {i}{48} e^{8A} G\star_6\bG\,,
\end{split}
\eqlabel{weq}
\end{equation}
\begin{equation}
\begin{split}
r^{(6)}_{mn}=&\ft 12 e^{-8A}\biggl(\tn_m e^{4A}\tn_n e^{4A}-16 
 \tn_m\ww\tn_n\ww\biggr)
-\LL\  e^{-4 A} \tg_{mn} +T_{mn}^{(1)}\\
&+
\ft 14 e^{4A} \biggl(G^{+}\ _{pqm}\bG^{-pq}\ _n+G^{-}\
_{pqm}\bG^{+pq}\  
_n
\biggr)\,,
\end{split}
\eqlabel{m6eq}
\end{equation}  
\begin{equation}
0=d\call+f^2\biggl(\bar{\call}\wedge d\calb+\ft 12 \call\wedge \left(
\calb d\bar{\calb}-\bar{\calb}d\calb\right)\biggr)\,,
\eqlabel{eql}
\end{equation}
\begin{equation}
f^2 \tn^2\calb+2f^4\bar{\calb}(\tn\calb)^2=-\ft{1}{12} e^{6A} G^+G^-\,.
\eqlabel{eqb}
\end{equation}
In \eqref{aeq}-\eqref{eqb} all index contractions are done with unwarped metric $\tg_{mn}$, 
$r^{(6)}_{mn}$ is the Ricci tensor constructed from $\tg_{mn}$, $\tn\equiv \n_y$,
$\star_6$ is defined on $\tcalm_6$,  
also 
\begin{equation}
\begin{split}
&G\bar{G}\equiv G_{mnp}\bar{G}^{mnp}\,, \qquad f^2\equiv \left(1-\calb\bar{\calb}\right)^{-1}\,,\\
&G^+\equiv \ft 12 G-\ft i2 \star_6 G\,,\qquad 
\qquad G^-\equiv \ft 12 G+\ft i2 \star_6 G\,,\\
&\call\equiv e^{4A} \star_6 G-4i\ww\ G\,,\\
&T_{mn}^{(1)}=\ft 14 \frac{\tn_m\tau\tn_n\bar{\tau}+
\tn_n\tau\tn_m\bar{\tau}}{(\Im{\tau})^2}\,.
\end{split}
\eqlabel{shorthands}
\end{equation} 
Notice that there is always solution to the 3-form Maxwell equation   \eqref{eql}, 
\begin{equation}
\call=0\qquad \Longleftrightarrow\qquad \star_6 G=4i e^{-4A} \ww\ G\,.
\eqlabel{l0}
\end{equation}
If all the following conditions are satisfied:  $\LL=0$, $\tcalm_6$ is a Calabi-Yau 3-fold,
$\calb=\rm const$, then $\ww=-\ft 14 e^{4A}$, and \eqref{l0} implies that the 
3-form flux $G$ is imaginary self-dual (ISD) \cite{ks,gkp}.   
We emphasize that while \eqref{l0} is always a solution,  it is not 
the most general solution. In fact,
supergravity backgrounds dual to four-dimensional  gauge theories with generic bare masses violate  
\eqref{l0}. 

A linear combination of \eqref{aeq} and \eqref{weq} give rise to 
\begin{equation}
\begin{split}
\tn^2\left( 4\ww+e^{4A}\right)=&e^{-4 A}\left(\tn\left[
4\ww+e^{4A}\right]\right)^2 +
{\textstyle\frac{1}{24}} e^{8A} |i G
+\star_6 G|^2+ 4\LL\,,\\
\tn^2\left( -4\ww+e^{4A}\right)=&e^{-4 A}\left(\tn\left[
-4\ww+e^{4A}\right]\right)^2 +
{\textstyle\frac{1}{24}} e^{8A} |i G
-\star_6 G|^2+ 4\LL\,.
\end{split}
\eqlabel{ddbpot}
\end{equation}
For the class of $\call=0$ solutions we further have
\begin{equation}
\begin{split}
\tn^2\left( 4\ww+e^{4A}\right)=&e^{-4 A}\left(\tn\left[
4\ww+e^{4A}\right]\right)^2 +
{\textstyle\frac{1}{24}} |G|^2 \left(4\ww+e^{4A}\right)^2
+ 4\LL\,,\\
\tn^2\left( -4\ww+e^{4A}\right)=&e^{-4 A}\left(\tn\left[
-4\ww+e^{4A}\right]\right)^2 +
{\textstyle\frac{1}{24}} |G|^2 \left(-4\ww+e^{4A}\right)^2+ 4\LL\,.
\end{split}
\eqlabel{ddbpotl0}
\end{equation}
The importance of \eqref{ddbpot} ( and \eqref{ddbpotl0}) stems from the fact that 
$\calv_{D3}$,  $\calv_{\bD3}$ defined according to 
\begin{equation}
\calv_{D3}\equiv T_3\left(-4\ww+e^{4A}\right)\,,\qquad \calv_{\bD3}\equiv T_3\left(4\ww+e^{4A}\right)\,,
\eqlabel{d3db3pote}
\end{equation} 
are precisely the potentials describing effective dynamics of $D3$ and $\bD3$ 
probe branes! 
Indeed, the effective action of a 3-brane probe of charge $q,\ |q|=1$ ($D3$ brane has $q=+1$ ), is   
\begin{equation}
S_{3}=-T_3\int_{\calm_4}d^4\xi\sqrt{-\hat{g}}+q T_3\int_{\calm_4}C^{(4)}\,,
\eqlabel{3brane}
\end{equation}
where $\hat{g}$ is the induced metric on the world-volume of the probe,
equal in the gauge $\xi^\mu=x^{\mu}$,\ \ie\ $d^4\xi\equiv d^4x$,  
\begin{equation}
\hat{g}_{\mu\nu}=e^{2A(y_q)}\ g_{\mu\nu}(x)+e^{-2A(y_q)}\ \tg_{mn}(y_q)\ \del_\mu y_q^m\ \del_{\nu} y_q^n\,,
\end{equation}
where $\{y_q^m=y_q^m(x)\}$ represents  the coordinates of the probe 3-brane in $\tcalm_6$.
Also, 
\begin{equation}
C^{(4)}=4\ww\ vol_{\calm_4}\,,
\eqlabel{c4}
\end{equation} 
where the factor of four comes from the different normalization of the four-form potential 
in \cite{schwarz83} and the one used in $Dp$-brane effective action \cite{pol2}. 
For slowly varying $y_q(x)$, we find an effective action
\begin{equation}
S_3=\int_{\calm_4} d^4x \sqrt{g}\biggl(-\ft 12 T_3\ \tg_{mn}(y_q)\ g^{\mu\nu}\ \del_\mu y_q^m\ \del_\nu y_q^n
-\calv_{q}(y_q)\biggr)\,,
\eqlabel{probeeff}
\end{equation}
where the effective potential $\calv_q$ is (compare with \eqref{d3db3pote})
\begin{equation}
\calv_q(y)=T_3\biggl(e^{4A(y)}-4q\ \ww(y)\biggr)\,.
\eqlabel{effpot}
\end{equation}
To extract a physical potential we need to rewrite it in terms of  canonical normalized  scalar fields $y_q^m(x)\to \phi^i(x)$,
\begin{equation}
\begin{split}
d\phi^i&\equiv T_3^{1/2} e^{i}_m(y)\ dy^m\,, \\
-\ft 12 T_3\ \tg_{mn}(y_q)\ g^{\mu\nu}\ \del_\mu y_q^m\ \del_\nu y_q^n&\Longrightarrow -\ft 12 \sum_{i=1}^6 
g^{\mu\nu}\ \del_\mu \phi^i\ \del_\nu \phi^i\,,
\end{split}
\eqlabel{gencan}
\end{equation}
where $e^i_m(y)$ are the vielbeins of the metric $\tg_{mn}(y)$. 
Finally, with \eqref{effpot} we can rewrite \eqref{ddbpot}, \eqref{ddbpotl0} as 
\begin{equation}
 \tn^2 \left(T_3^{-1}\calv_q\right)=e^{-4 A}\left(\tn\left[ T_3^{-1}\calv_q\right]\right)^2 +
{\textstyle\frac{1}{24}} e^{8A} |i G
-q\star_6 G|^2+ 4\LL\,,
\eqlabel{dpotg}
\end{equation}
for generic backgrounds, and for $\call=0$ backgrounds as 
\begin{equation}
 \tn^2 \left(T_3^{-1}\calv_q\right)=e^{-4 A}\left(\tn\left[ T_3^{-1}\calv_q\right]\right)^2 +
{\textstyle\frac{1}{24}} \left(T_3^{-1}\calv_q\right)^2 |G|^2+ 4\LL\,.
\eqlabel{dpotg0}
\end{equation}

\subsection{Effective mass of the radion (inflaton)}
In this section we study asymptotic behavior of a probe brane effective 
potential \eqref{dpotg}, \eqref{dpotg0} near the boundary of a warped 
type IIB supergravity background and determine the effective 
probe brane radion  mass. Our discussion is restricted to 
Euclidean geometries dual to mass deformed $\caln=4$ $SU(N)$ SYM theory on 
$\ss_4$ (or $S^4$), and to geometries dual to Klebanov-Strassler (KS) cascading gauge theories \cite{ks}     
on $\ss_4$ (or $S^4$ \cite{bo1}). 
In the former case, without any mass deformations, the $D3$ radion mass is that of a conformally coupled 
scalar $m_{conf}^2$
\begin{equation}
m_{rad}^2=m_{conf}^2\equiv\frac 23\ \LL\,.
\eqlabel{confm}
\end{equation}  
We find that turning on bare masses to fermionic components of the $\caln=4$ gauge theory 
chiral superfields (or appropriate 3-form fluxes in the dual supergravity 
background)  {\it always} raises the radion mass. On the other hand, turning on bare masses to bosonic components of the 
$\caln=4$ gauge theory chiral superfields (which corresponds to deforming the background geometry --- the round metric on $S^5$
in this case) can have either effect. These observations can be summarized as 
\begin{equation}
m_{rad}^2\ =\frac 23\ \LL\ +\ m^2_{fluxes}\ \pm\ m^2_{geometry}\,.
\eqlabel{mradn4}
\end{equation}
The last two terms in \eqref{mradn4} in principle can depend on the (squashed) $S^5$  angles,
in fact $m^2_{geometry}$ contribution might even change sign as a function of these angles. 
We will obtain an explicit expression for  \eqref{mradn4} in the case of $\caln=2^*$ 
PW flow on $\ss_4$ in section 4. 
For the gravitational dual to the deformed KS gauge theory we find
\begin{equation}
m_{rad}^2=\frac 23\ \LL\,,
\eqlabel{ksrad}
\end{equation}  
without any additional corrections. We should clarify that radion corrections 
from fluxes (and geometry deformation) are absent  if the three-form fluxes are induced 
by the fractional $D3$ branes {\it only} --- as in KS gauge theory gravitational dual. 
More general fluxes will lead to the modified radion mass as in \eqref{mradn4}.

Given \eqref{mradn4},  we see that for $\LL<0$, the tachyonic instability of the $D3$ radion is 
most efficiently confronted by giving mass only to fermionic components of the $\caln=4$ gauge theory chiral superfields.
Such a deformation necessarily completely breaks the supersymmetry. Even though supergravity dual 
to KS cascading gauge theory involves nontrivial fluxes, result \eqref{ksrad} implies that the 
$D3$ brane radion is tachyonic near the boundary. In some sense, the latter is expected, as prior to 
introducing the gauge theory background curvature, this gauge theory had a moduli space of vacua, and thus massless 
scalars. For the supergravity dual to KS gauge theory on $R^4$ \cite{ks}, these massless 
scalars are moduli of a $D3$ probe. Once the gauge theory background is deformed to a smooth 
quotient of $H_4$, $R^4\to \ss_4$, these scalars will develop a  mass, appropriate for a conformally coupled 
scalar \eqref{confm}. It must be  possible to give explicit bare mass to the KS moduli, thus removing 
the tachyons from the gauge theory spectrum on $\ss_4$. We did not attempt to construct corresponding deformations 
on the supergravity side.

Before we turn to the justification of above claims, it is instructive to see what \eqref{mradn4}, \eqref{ksrad}
imply for the {\it positive} four dimensional cosmological constant, $\LL>0$.  The reason why this is interesting 
for cosmological model building is discussed in \cite{br}. Briefly, gauge/string theory correspondence establishes 
an equivalence between a theory of dynamical gravity on direct warped product 
$\calm_4\times \tcalm_6$ and a non-gravitational  theory (gauge theory) on $\calm_4$. The non-gravitational 
feature of the effective theory on   $\calm_4$ is reflected in the non-compactness\footnote{
The non-compactness of $\tcalm_6$ is obviously a necessary condition for the dual boundary 
theory to be non-gravitational. It might very well be  that this condition is not sufficient.} of $\tcalm_6$ 
(the effective four dimensional Newton's constant vanishes). Compactifications of $\tcalm_6$ introduce 
dynamical gravity into low-energy effective four-dimensional picture \cite{rs,gkp}. 
Likewise, compactifications of the gravitational dual to gauge theory on de-Sitter space-time
\cite{bo1}, results in four-dimensional dynamical de-Sitter  vacua\footnote{Ref.~\cite{kklt} 
realizes a compactification of the gravitational dual of de-Sitter deformed KS cascading gauge theory. Embedding 
de-Sitter throats discussed in \cite{bo2,b2} into a global model is an open question.} \cite{kklt} (KKLT).        
Brane-anti-brane inflation in KKLT vacuum has been studied in \cite{kklt2} (K$^2$LM$^2$T). In  inflationary 
scenario of \cite{kklt2}, one has best computational control for a widely separated $D3\bD3$ pair, which is still 
deep inside (one of) the KS throat(s) of the global geometry. In this regime, inflaton can be identified 
with the radion of a $D3$ probe brane in the local (non-compact) geometry, dual to de-Sitter deformed cascading 
gauge theory \cite{br}. Thus,  \eqref{mradn4}, \eqref{ksrad} provide information about $\eta$-parameter 
of a single-field slow-roll brane inflation  of  K$^2$LM$^2$T  
\begin{equation}
\eta\equiv \frac{m_{rad}^2}{\LL}\,.
\eqlabel{etadef}
\end{equation}
Specifically, \eqref{ksrad} explains 'stability' of the anomalously large $\eta$-parameter observed in \cite{br}.
On the contrary, given \eqref{mradn4}, brane inflation in  de-Sitter throats constructed in \cite{b2}
can avoid this problem. Indeed, $m_{rad}^2$ can be made arbitrary small, without turning on any fluxes 
(fermionic mass terms), but fine tuning  masses of the bosonic components of the 
chiral superfields in the dual gauge theory language. Of cause, the latter requires 
the 'right sign' for the $m^2_{geometry}$ contribution. As we explicitly show  in section 4, this is 
straightforward to achieve.

\subsubsection{Mass deformed $\caln=4$ supergravity duals}
In this case the asymptotic\footnote{We keep only the leading terms.} metric on $\tcalm_6$ is flat
\begin{equation}
\begin{split}
\tg_{mn}(y)dy^mdy^n&\ \longrightarrow\ dr^2+r^2 (dS^5)^2\,,
\end{split}
\eqlabel{assmn4}
\end{equation}
where $r\to \infty$ is a radial coordinate, and $(dS^5)^2$ is the metric on a round $S^5$ . 
Additionally we have the following asymptotics for the warp factor $A(y)$ and the four-form 
potential $\ww(y)$ \eqref{c4}
\begin{equation}
\begin{split}
e^{A(y)}\ \longrightarrow\ \frac rL\,,\qquad \ww(y)\ \longrightarrow\ \frac {r^4}{4L^4}\,,\qquad r\to \infty\,.
\end{split}
\eqlabel{n4r1}
\end{equation}
Finally, following the gauge/string theory correspondence dictionary \cite{adscft}, 
component of the three-form fluxes $G_{I_1I_2I_3}$ corresponding to masses of the fermionic components of the 
dual $\caln=4$ gauge 
theory  chiral superfields, in the orthonormal frame of  \eqref{10metricgg}, scale near the boundary  as
\begin{equation}
G_{I_1I_2I_3}\ \sim\ \frac 1r\,,
\eqlabel{n4r10}
\end{equation}
which corresponds to 
\begin{equation}
|iG-q\star_6 G|^2=(iG-q \star_6 G)_{mnp}(-i\bar{G}-q \star_6\bar{G})^{mnp}\ \longrightarrow\ 
\frac{L^8}{r^8}\ \calg^2_q\,,\qquad r\to\infty\,,
\eqlabel{n4r11}
\end{equation}  
where $\calg^2_q\equiv \calg^2_q\left(\om_{S^5}\right)$ is a non-negative function of the $S^5$ angles, 
detailed form of which depends on the fermionic mass matrix.
As before, we identify the scalar in the effective 
$D3$ probe brane action associated with its motion in $r$ direction with the radion. 
Then, using \eqref{gencan} and the asymptotic form of the metric \eqref{assmn4} we conclude 
\begin{equation}
\phi\ \longrightarrow\ T_3^{1/3} r\,,\qquad r\to \infty\,,
\eqlabel{n4r0}
\end{equation}
which results in 
\begin{equation}
\begin{split}
\tn^2&\ \longrightarrow\ T_{3}\left[\ \frac{\del^2}{\del\phi^2}+\frac{5}{\phi}\ \frac{\del}{\del\phi}+\frac{1}{\phi^2}\ 
\n^2_{S^5}\ \right]\,,\\
\tn&\  \longrightarrow\ T^{1/2}_3\left\{\frac{\del}{\del\phi},\frac{1}{\phi}\n_{S^5}\right\}\,,
\end{split}
\eqlabel{n4r2}
\end{equation}
as  $\phi\to\infty$. 
In \eqref{n4r2}, $\n_{S^5}^2$ is a Laplacian on a round $S^5$. 
Notice that with \eqref{n4r1}, the coefficient of the leading scaling $(\sim r^4)$ of the effective 
$D3$ probe brane potential $\calv_{D3}$ \eqref{d3db3pote} near the boundary vanishes. Thus we expect asymptotically 
as $r\to \infty$ (or $\phi\to \infty$)
\begin{equation}
\calv_{D3}\equiv \calv_{rad}=\frac 12\ m_{rad}^2\left(\om_{S^5}\right)\ \phi^2+\calo(\phi^0)\,,
\eqlabel{n4r3}
\end{equation}
where we explicitly indicated potential dependence of $m_{rad}^2$ on the $S^5$ angles.
Given the asymptotics \eqref{assmn4}-\eqref{n4r3}, we find from \eqref{dpotg}
\begin{equation}
m_{rad}^2=\frac 23\ \LL+\frac{1}{144}\ \calg^2_{+1}-\frac{1}{12}\ \n^2_{S^5}\left(m_{rad}^2\right)+\calo(\phi^{-2})\,,
\eqlabel{n4fin}
\end{equation}
resulting in \eqref{mradn4} with the identifications
\begin{equation}
m^2_{fluxes}\equiv \frac{1}{144}\ \calg^2_{+1}\,,\qquad \pm m^2_{geometry}\equiv-\frac{1}{12}\ \n^2_{S^5}\left(m_{rad}^2\right)\,,
\eqlabel{idn4}
\end{equation}
where the $\pm$ is to indicated that  $\n^2_{S^5}\left(m_{rad}^2\right)$ can change sign on the $S^5$. 
We will see an explicit example of this in section 4.

\subsubsection{KS supergravity duals}
In this case the analysis is slightly different. 
All the asymptotics can be extracted from the Klebanov-Tseytlin (KT) solution \cite{kt}.  
As before, asymptotic  metric on $\tcalm_6$ is flat
\begin{equation}
\begin{split}
\tg_{mn}(y)dy^mdy^n&\ \longrightarrow\ dr^2+r^2 (dT^{1,1})^2\,,\\
\end{split}
\eqlabel{assmks}
\end{equation}
where $r\to \infty$ is a radial coordinate, $(dT^{1,1})^2$ is the metric on the angular part 
of the six-dimensional conifold, $T^{1,1}\equiv \frac{SU(2)\times SU(2)}{U(1)}$. 
Additionally we have the following asymptotics for the warp factor $A(y)$ and the four-form 
potential $\ww(y)$ \eqref{c4}
\begin{equation}
\begin{split}
e^{A(y)}=e^{A(r)}\ \longrightarrow\ \frac {r}{L\ln^{1/4}r}\,,\qquad \ww(y)=\ww(r)\ 
\longrightarrow\ \frac {r^4}{4L^4\ln r}\,,\qquad r\to \infty\,.
\end{split}
\eqlabel{ksr1}
\end{equation}
Notice that there is no dependence on $T^{1,1}$ coordinates for $A(y),\ \ww(y)$.
This immediately implies that the effective probe brane potential $\calv_q$ \eqref{effpot}
is a function of $r$ only.

It is possible to extract the scaling of the three-form flux directly from \cite{kt} 
(or corresponding deformed solution \cite{bo1}). Here, we motivate the answer.  
In mass deformed $\caln=4$ supergravity duals the RG flow is induced by three-form 
fluxes dual to these masses. In the KS solution, the RG flow is induced by the three-form flux from 
fractional $D3$-branes ($D5$ branes wrapping a 2-cycle of the conifold). The $F_3$ flux though the 3-cycle of the 
conifold (transverse to  $D5$ branes ) is topological, thus given \eqref{assmks}, 
$F_3^2\equiv F_{3\ mnp}F_3^{mnp}\sim r^{-6}$. Altogether, we find
\begin{equation}
|G|^2=G_{mnp}\bar{G}^{mnp}\ \longrightarrow\ 
\frac{L^6}{r^6}\ \calg^2\,,\qquad r\to\infty\,,
\eqlabel{ksr11}
\end{equation}  
where $\calg^2\equiv \calg^2\left(\om_{T^{1,1}}\right)$ is a non-negative function of the $T^{1,1}$ angles.
It's precise form is not important in what follows.
Again, we identify the scalar in the effective 
$D3$ probe brane action associated with motion in $r$ direction with the radion. 
Using \eqref{gencan} and the asymptotic form of the metric \eqref{assmks} we conclude 
\begin{equation}
\phi\ \longrightarrow\ T_3^{1/3} r\,,\qquad r\to \infty\,,
\eqlabel{ksr0}
\end{equation}
which results in 
\begin{equation}
\begin{split}
\tn^2&\ \longrightarrow\ T_{3}\left[\ \frac{\del^2}{\del\phi^2}+\frac{5}{\phi}\ \frac{\del}{\del\phi}+\frac{1}{\phi^2}\ 
\n^2_{T^{1,1}}\ \right]\,,\\
\tn&\  \longrightarrow\ T^{1/2}_3\left\{\frac{\del}{\del\phi},\frac{1}{\phi}\n_{T^{1,1}}\right\}\,,
\end{split}
\eqlabel{ksr2}
\end{equation}
as  $\phi\to\infty$. 
In \eqref{ksr2}, $\n_{T^{1,1}}^2$ is a Laplacian on  $T^{1,1}$. 
As before,  with \eqref{ksr1}, the coefficient of the leading scaling $(\sim r^4)$ of the effective 
$D3$ probe brane potential $\calv_{D3}$ \eqref{d3db3pote} near the boundary vanishes. Thus we expect asymptotically 
as $r\to \infty$ (or $\phi\to \infty$)
\begin{equation}
\calv_{D3}\equiv \calv_{rad}=\frac 12\ m_{rad}^2\ \phi^2+\calo(\phi^0)\,,
\eqlabel{ksr3}
\end{equation}
though without any  dependence of $m_{rad}^2$ on the $T^{1,1}$ 
angles.
It is crucial that as for the original KT/KS solution, the three-form 
fluxes for their $\LL\ne 0$  deformations solve Maxwell equations with $\call=0$, 
\eqref{l0}. Thus, with  the asymptotics \eqref{assmks}-\eqref{ksr3}, we find from \eqref{dpotg0}
\begin{equation}
m_{rad}^2=\frac 23\ \LL\,,
\eqlabel{ksfin}
\end{equation}
resulting in \eqref{ksrad}.
The same conclusion can be reached  for the  more 
general ansatz for $\calv_{rad}$, $\calv_{rad}\sim \phi^2\ \ln^n\phi$ as $\phi\to\infty$.

\section{$\caln=2^*$ flow on $\ss_4$}  
Here we consider the $\caln=2^*$ Pilch-Warner flow\footnote{The dual gauge theory picture for the 
PW supergravity flow is explained in \cite{bpp,j}.} 
\cite{pw} on smooth compact quotients of Euclidean $AdS_4$, or $H_4$.
 Closely related deformations  of this RG flow where discussed in \cite{b2}. 
We present a complete ten-dimensional non-supersymmetric solution of type IIB supergravity
realizing this flow, and study the $D3$ probe brane dynamics in this background. In agreement with general arguments of the 
previous section, we find that the probe brane instabilities can be lifted once sufficiently 
large three-form flux  corresponding to masses of the $\caln=2$ hypermultiplet fermionic components 
are turned on. Supergravity background metric deformations dual to turning masses for the  bosonic components of the 
$\caln=2$ hypermultiplet contribute to the radion mass as explained in section 3.2. 
For zero masses of the hypermultiplet components, the supergravity solution is a
Euclidean wormhole recently studied in \cite{ml}. We determine (analytically) deformation of this wormhole 
solution induced by small hypermultiplet masses. We then study numerically the deformed 
wormhole solution as we increase the fermionic mass parameter. We find that before the
radion of the $D3$ probe (for $\LL<0$) ceases to be tachyonic, the background geometry 
develops a naked singularity. 
Though we presented an explicit scenario where a physically well-motivated 
stabilization of the wormhole instability fails, it is a bit premature to claim 
that a smooth, single-boundary 
solution, free from the non-perturbative instabilities due to $D3\bD3$ production, in this model
does not exist. Such a claim would require an understanding of the resolution of the 
naked time-like singularity in the model for large fermionic mass parameters. 
We hope to return to this problem in the future.

In conclusion, we observe that it might be possible to obtain slow-roll brane inflation in 
de-Sitter deformed ($\LL>0$) $\caln=2^*$ throat geometries \cite{bg}.

\subsection{Background and the $D3$ probe dynamics}

We begin the background construction in five-dimensional supergravity, and will further 
uplift the solution to ten dimensions. 
The effective 5d action is 
\begin{equation}
S=\int d\xi^5 \sqrt{-g}\left(\frac 14  R-3(\del\a)^2-(\del\chi)^2-
\calp\right)\,,
\eqlabel{action5}
\end{equation} 
where the potential $\calp$ is\footnote{We set the 5d gauged SUGRA coupling 
to one. This corresponds to setting $S^5$ radius $L=2$.} 
\begin{equation}
\calp=\frac{1}{48} \left(\frac{\del W}{\del \a}\right)^2+
\frac{1}{16} \left(\frac{\del W}{\del \chi}\right)^2-\frac 13 W^2\,,
\eqlabel{pp}
\end{equation}
with the superpotential
\begin{equation}
W=-\frac{1}{\r^2}-\frac 12 \r^4 \cosh(2\chi)\,.
\eqlabel{supp}
\end{equation}
From \eqref{action5} we have Einstein equations
\begin{equation}
\frac 14 R_{\mu\nu}=3\del_\mu \a \del_\nu\a+\del_\mu\chi\del_\nu\chi
+\frac 
13 g_{\mu\nu} \calp\,,
\eqlabel{ee}
\end{equation} 
plus the scalar equations
\begin{equation}
\begin{split}
0&=\frac {6}{\sqrt{-g}}\del_\mu \left(g^{\mu\nu} \sqrt{-g}\ \del_\mu\a \right)
-\frac{\del\calp}{\del \a}\,,\\
0&=\frac {2}{\sqrt{-g}}\del_\mu \left(g^{\mu\nu} \sqrt{-g}\ \del_\mu\chi\right)
-\frac{\del\calp}{\del \chi}\,.\\
\end{split}
\eqlabel{scalar}
\end{equation}
With the RG flow metric
\begin{equation}
\begin{split}
\qquad ds_5^2&=e^{2 A} ds_{\ss_4}^2+dr^2\,,\\
\end{split}
\eqlabel{ab}
\end{equation}
the equations of motion \eqref{ee}, \eqref{scalar} become
\begin{equation}
\begin{split}
0&=\a''+4 A'\a' -\frac 16 \frac{\del\calp}{\del\a}\,,\\
0&=\chi''+4 A'\chi' -\frac 12 \frac{\del\calp}{\del\chi}\,,\\
&\frac 14 A''+\left(A'\right)^2+\frac 34 e^{-2 A}=-\frac 13 \calp\,,\\
&- A''-\left(A'\right)^2
=3\left(\a'\right)^2+\left(\chi'\right)^2 +\frac 13 \calp\,.
\end{split}
\eqlabel{beq}
\end{equation}
Though we can not find solution to \eqref{beq} analytically, it is straightforward 
to construct asymptotic solution as $r\to\infty$.
To analyze the  ultraviolet ($r\to\infty$) asymptotics it is convenient to introduce a new radial coordinate 
\begin{equation}
x\equiv e^{-r/2}\,.
\end{equation}
We find 
\begin{equation}
\begin{split}
A=&\xi-\ln x+x^2\biggl(e^{-2\xi}-\ft 13 \chi_0^2\biggr)+x^4\biggl(\ft 19\chi_0^4
-\ft 12 e^{-4\xi}-\ft 16 \chi_0^2 e^{-2\xi}-\ft 12 \chi_0^2\chi_{10}-\r_{10}^2-\ft 18\r_{11}^2
\\ &+\left(2\chi_0^2e^{-2\xi}-\ft 23 \chi_0^4-2\r_{10}\r_{11}
\right)\ \ln x-\r_{11}\ \ln^2 x\biggr)+\calo(x^6\ \ln^3 x)\,,
\end{split}
\eqlabel{uvseries1}
\end{equation}
\begin{equation}
\begin{split}
\r=&1+x^2\biggl(\r_{10}+\r_{11}\ \ln x\biggr)+x^4\biggl(
\ft 13 \chi_0^4+\ft 32 \r_{10}^2-2\r_{10}\r_{11}+\ft 32 \r_{11}^2+\ft 23 \chi_0^2(5\r_{10}-4\r_{11})\\&-2e^{-2\xi}(2\r_{10}-\r_{11})
+\left(\ft{10}{3}\chi_0^2\r_{11}+3\r_{10}\r_{11}-2\r_{11}^2-4\r_{11} e^{-2\xi}\right)\ \ln x\\
&+\ft 32 \r_{11}^2\ \ln^2 x\biggr)+\calo(x^6\ \ln^3 x)\,, 
\end{split}
\eqlabel{uvseries2}
\end{equation}
\begin{equation}
\begin{split}
\chi=&\chi_0 x \biggl(
1+x^2\biggl(\chi_{10}+\left(\ft 43 \chi_0^2-4e^{-2\xi}\right)\ \ln x\biggr)
\biggr)+\calo(x^5\ \ln^2 x)\,,
\end{split}
\eqlabel{uvseries3}
\end{equation}
where $\{\xi,\chi_0,\chi_{10},\r_{10},\r_{11}\}$ are parameters characterizing the asymptotics.
As explained in \cite{bl}, $\r_{11}$ ($\chi_0$) should be identified with the mass $m^2_{b}$ ($m_{f}$) of the bosonic (fermionic) components 
of the $\caln=2$ hypermultiplet.  Two more parameters $\r_{10},\ \chi_{10}$ are related to the bosonic and fermionic 
bilinear condensates correspondingly. Finally, $\xi$ is a residual  integration constant associated with 
fixing the radial coordinate --- it can be removed at the expense of shifting the origin of the radial coordinate $r$, 
or rescaling $x$.

The complete ten-dimensional lift of the RG flow \eqref{beq} is given in the Appendix. 
As in section 2.2, we consider a $D3$ probe slowly moving along  the radial direction.
Using the general expression \eqref{effpot} (with $q=+1$), and explicit ten-dimensional flow expressions
\eqref{10m}, \eqref{wres1} we find
\begin{equation}
\calv=T_3\left(\om^4 e^{4A}-4\ww\right)\,.
\eqlabel{potPW} 
\end{equation}
For the canonically normalized radion field  $\phi_r$ we find using   \eqref{10m}
\begin{equation}
d\phi_r=T_3^{1/2} e^A \om^2\  dr\,.
\eqlabel{canrad}
\end{equation} 
Now the mass of the radion close to the boundary is given 
\begin{equation}
\begin{split}
m_{\phi_r}^2=&\lim_{\phi_r\to\infty}\ \frac{\del^2}{\del\phi_r^2}\ \calv\\
=&\lim_{r\to\infty}\ \om^{-2} e^{-A}\ \frac{\del}{\del r} \biggl[
 \om^{-2} e^{-A}\ \frac{\del}{\del r} \calv
\biggr]\,.
\end{split}
\eqlabel{massR}
\end{equation} 
Using the asymptotics \eqref{uvseries1}-\eqref{uvseries3} we find 
\begin{equation}
\begin{split}
m_{\phi_r}^2=&-2+\biggl[\ft 23 e^{2\xi} \chi_0^2\biggr]
+\biggl[e^{2\xi}\r_{11}\left(\ft 32 \cos^2\theta-1\right)\biggr]\,.
\end{split}
\eqlabel{m2res}
\end{equation}  
Eq.~\eqref{m2res} should be compared with \eqref{mradn4} (here $\LL=-3$). 
Given explicit expressions for the three-form fluxes and the ten-dimensional lift of the background geometry 
\eqref{a2}, \eqref{10m} we can also verify identifications \eqref{idn4}.
From \eqref{m2res} we see that the instabilities will go away (the radion mass is always positive)
provided 
\begin{equation}
e^{2\xi}\left(\ft 23 \chi_0^2-\rho_{11}\right)-2\ge 0\,.
\eqlabel{rinst}
\end{equation}
Implicit in the result \eqref{rinst} went the condition $\r_{11}>0$, which is indeed the case 
as $\r_{11}$ is related to the  mass square of the bosonic components of the chiral superfields 
inducing $\caln=2^*$ RG flow.  Interestingly, the asymptotic $\caln=2^*$ supersymmetry 
requires $\ft 23 \chi_0^2=\rho_{11}$ \cite{bl}, for which we will always have instabilities.

\subsection{Analytical wormhole solution for small masses}
Stability analysis of the previous section rely only on the boundary behavior of the 
geometry and fluxes. It is important to establish whether  mass deformed RG flows for the 
$\caln=4$ $SU(N)$ SYM on $\ss_4$ are singularity free in the infrared. Here we show that this is 
indeed the case at least for small masses.

First of all, we have a Euclidean wormhole solution
\begin{equation}
\begin{split}
A=&\ln\left(\cosh\frac r2\right)\,,\\
\r=&1\,,\qquad \chi=0\,,
\end{split}
\eqlabel{zeroorder}
\end{equation}
which corresponds to turning off all the masses. This is just Euclidean $AdS_5/\Gamma$ written
in hyperbolic $\ss_4=H_4/\Gamma$ slicing.   
We will now construct leading order in mass parameters deformation of the wormhole \eqref{zeroorder}. 
Specifically,  we look for the solution to leading order in 
$\a_1,\a_2$  to \eqref{beq} (a similar recipe is employed in \cite{bl}) within the ansatz
\begin{equation}
\begin{split}
A=&\ln\left(\cosh\frac r2\right)+\a_1^2\ a_1(r)+\a_2^2\ a_2(r)\,,\\
\chi=&\a_1\ \chi_1(r)\,,\\
\r=&\a_2\ \r_2(r)\,.
\end{split}
\eqlabel{smass}
\end{equation}
We find 
\begin{equation}
\begin{split}
\chi_1=&\dd_1\ \frac{1}{\cosh^3\ft r2}+\dd_2\ \frac{\sinh r+r}{\cosh^3\ft r2}\,,\\
\r_2=&\dd_3\ \frac{\sinh\ft r2}{\cosh^3\ft r2}+\dd_4\ \frac{r \sinh \ft r2-2\cosh\ft r2}{\cosh^3\ft r2}\,,
\end{split}
\eqlabel{chi1res}
\end{equation}
where $\dd_i$ are integration constants. 
Additionally we have 
\begin{equation}
\begin{split}
0=&\sinh r\  a_1'-\cosh^2\ft r2 \left(\ft 12\chi_1^2+\ft 23\left(\chi_1'\right)^2\right)-a_1\,,\\
0=&\sinh r\  a_2'-2\cosh^2\ft r2 \left(\r_2^2+\left(\r_2'\right)^2\right)-a_2\,.
\end{split}
\eqlabel{aidiff}
\end{equation}
Both equations in \eqref{aidiff} can be analytically integrated, though result is not illuminating. What is important is 
that the deformation of the wormhole solution \eqref{zeroorder} by small 'bosonic' and 'fermionic' mass 
parameters exist --- it is still a wormhole.

\subsection{Wormhole solution without instability?}

\begin{figure}[t]
\begin{center}
\epsfig{file=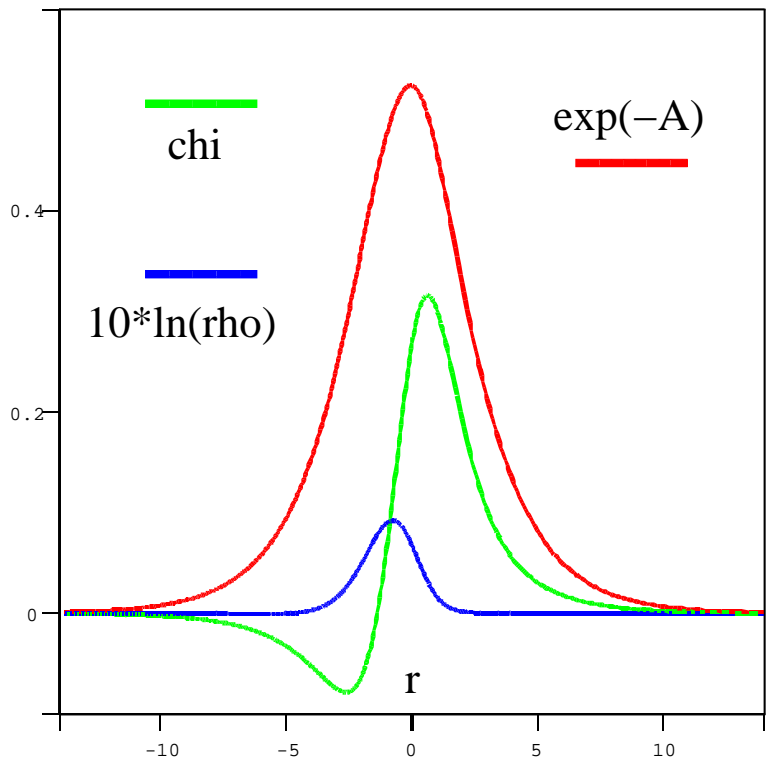,width=0.7\textwidth}
\caption{
For  small fermionic mass parameter $\chi_0$ the Euclidean wormhole solution 
(albeit non-perturbatively unstable)  is singularity-free. 
A typical RG flow with $|\chi_0|<\chi_{singular}$. Here $\chi_0=\ft 15 \chi_{critical}$. }
\label{cases}
\end{center}
\end{figure}

\begin{figure}[t]
\begin{center}
\epsfig{file=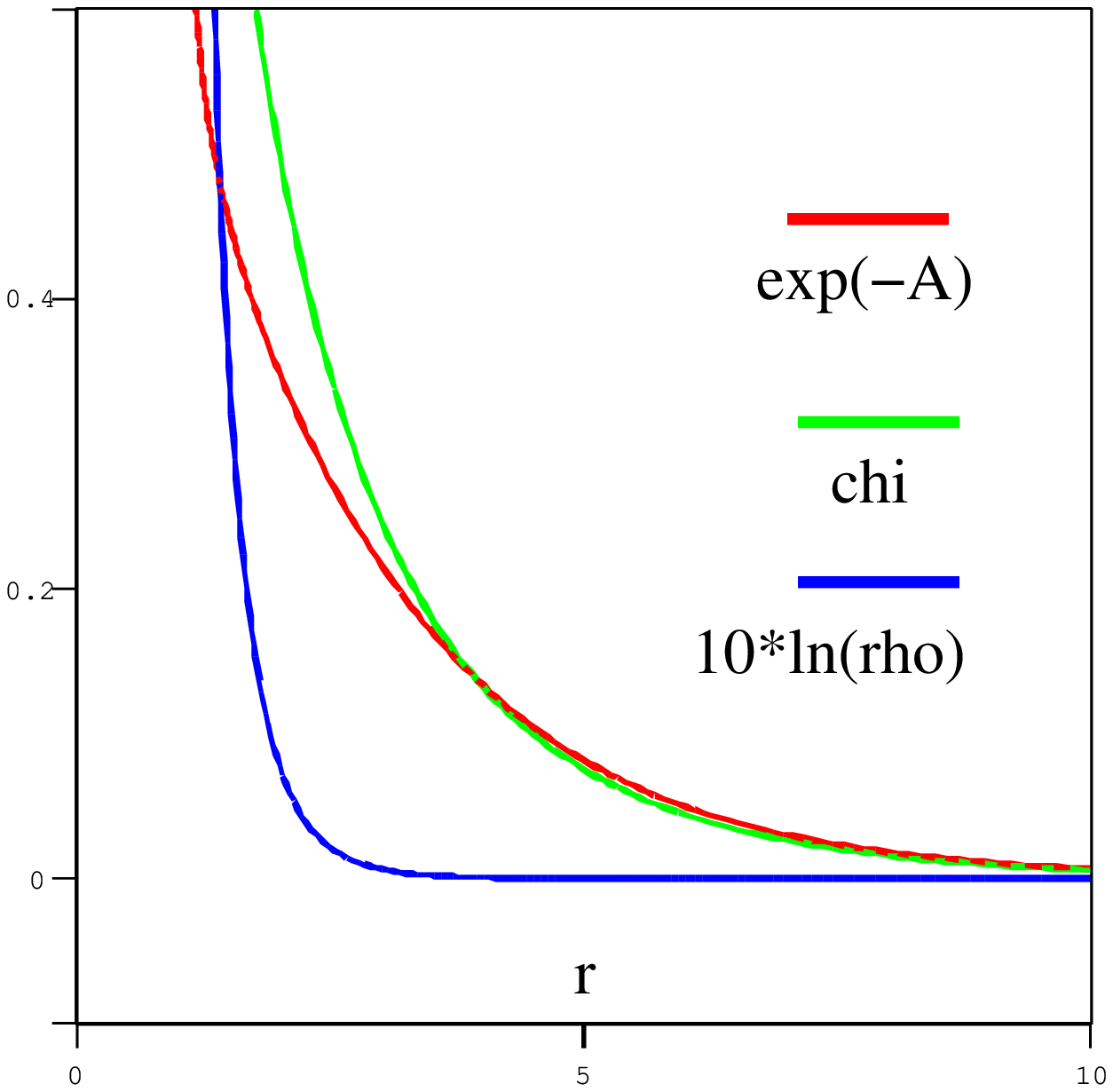,width=0.7\textwidth}
\caption{
As we increase the  fermionic mass parameter $|\chi_0|$, the Euclidean wormhole solution 
develops a naked singularity, before it can be stabilized against $D3\bD3$ pair 
production.  
A typical RG flow with $\chi_{critical}>|\chi_0|>\chi_{singular}$. Here $\chi_0=\ft 12 \chi_{critical}$. }
\label{cases1}
\end{center}
\end{figure}

In previous section we demonstrated that the wormhole \eqref{zeroorder} persists for small 
deformations, corresponding to turning on masses. From \eqref{rinst} small mass
can't cure the Schwinger-like instabilities of wormhole geometries due to the 
$D3\bD3$ pair production.  
We would like to ask the question what happens with the wormhole solution once this instability 
is removed. To simplify the problem we will turn on only the fermionic 
masses\footnote{
Though further numerical analysis are desirable, we do not believe that they will change the qualitative picture 
that emerges here.
}, \ie, we set $\r_{11}=\r_{10}=\chi_{10}=0$. 
Inspection of RG flow equations shows that $\r(r)$ is still a nontrivial 
function. This is just a reflection of the fact that bosonic masses are induced by higher loop effects, 
even though bare  masses are set to zero.
Without loss of generality we can choose the radial coordinate in such a way that $\xi=0$. 
Then 
\eqref{rinst} translates into 
\begin{equation}
\chi_0^2>3\equiv \chi_{critical}^2\,.
\eqlabel{ccritical}
\end{equation}
Numerical integration of \eqref{beq}, with boundary data dependent only on $\chi_0$ as outlined above, 
reveals two different types of RG flows, separated by $\chi_{singular}$,
\begin{equation}
\frac{\chi_{singular}}{\chi_{critical}}\approx 0.3719\cdots\,.
\eqlabel{ccdef}
\end{equation}
For 
\begin{equation}
|\chi_0|< \chi_{singular}\,,
\end{equation}
the RG flow geometry is a smooth (albeit non-perturbatively unstable) wormhole. 
A typical behavior of the warp factor $A(r)$, and the 5d gauged supergravity scalars $\chi(r),\ \rho(r)$
is shown in Fig.~1.

As the fermionic mass parameter $|\chi_0|$ is increased above $\chi_{singular}$
\eqref{ccdef}, the background geometry develops a naked singularity. This singularity
is associated with collapsing to zero size $\ss_4$, and correspondingly with  the divergence of the 
stress-tensor of the supergravity scalars $\chi$ and $\r$. A typical behavior of the 
RG flow in this regime is shown in Fig.~2. Since $\chi_{singular}<\chi_{critical}$,
the Euclidean wormhole solution develops a naked singularity before it can be stabilized.

\subsection{A comment on slow-role inflation in de-Sitter deformed $\caln=2^*$ throats}
One of the problems of brane inflation is generically
large $\eta$ parameter \eqref{etadef}, \cite{kklt2,br}.  
We argue here that it appears to be  possible to achieve slow-roll brane inflation in de-Sitter
deformed $\caln=2^*$ throats, $\LL>0$. Specifically, we demonstrate that $\eta$ can be made 
arbitrary small. Detailed study of this cosmological model will appear elsewhere \cite{bg}.

Reintroducing $\LL$, \eqref{m2res} becomes
\begin{equation}
m_{\phi_r}^2=\frac 23\ \LL+\biggl[\ft 23 e^{2\xi} \chi_0^2\biggr]
+\biggl[e^{2\xi}\r_{11}\left(\ft 32 \cos^2\theta-1\right)\biggr]\,,
\eqlabel{m2resp}
\end{equation}
thus leading to 
\begin{equation}
\eta=\frac 23 +\biggl[\ft 23 \LL^{-1}\ e^{2\xi} \chi_0^2\biggr]
+\biggl[\LL^{-1}\ e^{2\xi}\r_{11}\left(\ft 32 \cos^2\theta-1\right)\biggr]\,.
\eqlabel{ef}
\end{equation}
We see that to reduce $\eta$, we, first of all, would like to 
turn off 3-form fluxes (fermionic mass parameter), \ie, set $\chi_0=0$. 
In fact, setting $\chi(r)\equiv 0$ is a consistent truncation of the 
full RG flow equations, \eqref{beq}. From \eqref{m2resp}, it is clear 
that  a $D3$ probe would tend to move in the $\cos\theta=0$ 'valley', where its 
potential energy is locally minimized\footnote{For $\LL=0$ this submanifold is a 
moduli space of a $D3$ probe in the PW background \cite{bpp,j}.}.   
If we now identify the effective inflaton field with the radial motion of the 
$D3$ probe in the $\cos\theta=0$ valley,  its $\eta$ parameter becomes
\begin{equation}
\eta=\frac 23-\LL^{-1}\ e^{2\xi}\r_{11}\,,
\eqlabel{eff}
\end{equation}
which can be made arbitrary small by fine-tuning the deformation parameter $\r_{11}$,
corresponding to turning on masses to bosonic components of the $\caln=2$ hypermultiplet,
$m_b^2\sim \LL$. Given general arguments of \cite{bt}, we expect such backgrounds to be 
singularity-free.


\section*{Acknowledgments}
I would like  to thank Vic Elias, Chris Herzog, Gerry McKeon, Volodya Miransky and  Rob Myers
for valuable discussions. 
Research at the Perimeter Institute is supported in part by funds from NSERC of 
Canada.

\section*{Appendix}
Here we present ten-dimensional  lift of five-dimensional RG flow of section 4.1. 
The 10d Einstein frame metric is 
\begin{equation}
\begin{split}
ds_{10}^2&=\Omega^2 ds_5^2 + 4 \frac {(c X_1 X_2)^{1/4} }{\r^3}\biggl(
c^{-1} d\theta^2+\r^6\cos^2\theta \left(\frac {\sigma_1^2}{c X_2}
+\frac{\sigma_2^2+\sigma_3^2}{X_1}\right)+\sin^2\theta\frac {d\phi^2}{X_2}
\biggr)\,,
\end{split}
\eqlabel{10m}
\end{equation}  
where $ds_5^2$ is the five-dimensional  flow metric \eqref{ab},  
$c\equiv \cosh (2\chi)$. The warp factor is given by
\begin{equation}
\Omega^2=\frac {(c X_1 X_2)^{1/4} }{\r}\,,
\eqlabel{om5}
\end{equation}
and the two functions $X_i$ are defined by
\begin{equation}
\begin{split}
X_1(r,\theta)&=\cos^2\theta+\r(r)^6\cosh(2\chi(r))\sin^2\theta\,,\\
X_2(r,\theta)&=\cosh(2\chi(r))\cos^2\theta+\r(r)^6\sin^2\theta\,.
\end{split}
\eqlabel{x1x2}
\end{equation}
Additionally, $\sigma_i$ are the $SU(2)$ left-invariant forms normalized so that 
$d\sigma_i=2 \sigma_j\wedge \sigma_k$.
For the dilaton/axion (compare with \eqref{axidil}, \eqref{shorthands}) we have 
\begin{equation}
f=\frac 12 \left(\left(\frac{c X_1 }{X_2}\right)^{1/4}+
\left(\frac{c X_1 }{X_2}\right)^{-1/4}\right),\qquad 
f\calb =\frac 12 \left(\left(\frac{c X_1 }{X_2}\right)^{1/4}-
\left(\frac{c X_1 }{X_2}\right)^{-1/4}\right) e^{2i \phi}\,.
\eqlabel{dilax}
\end{equation} 
The 3-form fluxes are 
\begin{equation}
A_{(2)}=e^{i\phi}\left(a_1(r,\theta)\ d\theta\wedge \sigma_1+a_2(r,\theta)\
 \sigma_2
\wedge \sigma_3+a_3(r,\theta)\ \sigma_1\wedge d\phi+a_4(r,\theta)\ 
d\theta\wedge d\phi\right)\,,
\eqlabel{a2}  
\end{equation}
where $a_i(r,\theta)$ are given by 
\begin{equation}
\begin{split}
a_1&=- i\ 4\ \tanh(2\chi) \cos\theta\,, \\
a_2&=i\ 4\ \frac{\r^6\sinh(2\chi)}{X_1}\ \sin\theta\cos^2\theta\,,\\
a_3&=-4\  \frac{\sinh(2\chi)}{X_2}\ \sin\theta\cos^2\theta\,,\\
a_4&=0\,.
\end{split}
\eqlabel{aaa1}
\end{equation}
Finally, the 5-form flux is
\begin{equation}
\begin{split}
\qquad F_5=\calf+\star\calf\,,\qquad 
\calf={\rm vol}_{H^4}\wedge d\omega\,,
\end{split}
\eqlabel{5form5}
\end{equation}
where $\omega(r,\theta)$ satisfies 
 \begin{equation}
\begin{split}
&\frac{\del \omega}{\del \theta}=-\frac 32 e^{4 A } \left(\ln \r\right)'\ 
\sin 2\theta\,,\\
&\frac{\del \omega}{\del r}=\frac 18 e^{4 A }\ 
\frac{1}{\r^4}\ \biggl(-\r^{12}\sinh^2(2\chi)\sin^2\theta +2\r^6 
\cosh(2\chi)(1+\sin^2\theta)+2\cos^2\theta\biggr) \,.
\end{split}
\eqlabel{wres1}
\end{equation}

\end{document}